\documentclass{article}
\usepackage{spconf,amsmath,epsfig,amsfonts}
\usepackage{psfrag}
\usepackage{amsmath}
\usepackage{amssymb}
\usepackage{amsfonts}
\usepackage{color}

\def\PSNR{\mathrm{ PSNR}}
\def\punit{\, \mathrm}

\title{Spatio-temporal prediction in video coding by spatially refined motion compensation}
\name{J\"urgen~Seiler and Andr\'e~Kaup}
\address{Chair of Multimedia Communications and Signal Processing, \\University of Erlangen-Nuremberg, Cauerstr. 7, 91058 Erlangen, Germany\\
{\{seiler, kaup\}@LNT.de}}

\begin{document}
\topmargin=0mm
\ninept
\maketitle


\begin{abstract} \label{abstract}
The purpose of this contribution is to introduce a new method of signal prediction in video coding. Unlike most existent prediction methods that either use temporal or use spatial correlations to generate the prediction signal,  the proposed method uses spatial and temporal correlations at the same time. The spatio-temporal prediction is obtained by first performing motion compensation for a macroblock, followed by a refinement step that pays attention to the correlations between the macroblock and its surroundings. At the decoder, the refinement step can be performed in the same manner, thus no additional side information has to be transmitted. Implementation of the spatial refinement step into the H.264/AVC video codec leads to reduction in data rate of up to nearly 15\% and increase in PSNR of up to 0.75 dB, compared to pure motion compensated prediction.
\end{abstract}


\begin{keywords}
Spatio-Temporal Prediction, Video Coding, Video Communication
\end{keywords}


\section{Introduction} \label{sec:introduction}

The transmission and playback of video signals has become still more and more important in many modern communication systems. This is only possible since video data can be strongly compressed. One major cause for being able to compress video data is the prediction of the video signal during the encoding. Thereby, for the area actually being coded, the encoder aims at generating an estimate for the video signal based on already transmitted parts of the signal. These parts either can be previous frames or regions of the actual frame that already have been processed. As the signal in these regions is also known to the decoder, the decoder can perform the signal prediction equivalently to the encoder. Hence, not the quantized and entropy coded video signal itself, but only the quantized and entropy coded difference signal between the original video signal and the predicted signal has to be transmitted. Consequently, the compression efficiency of a video codec strongly depends on its ability to predict the signal being coded.

Modern video codecs, as e.\ g.\ the H.264/AVC \cite{Ostermann2004}, use either temporal or spatial correlations for predicting the video signal. Normally, the spatial prediction is performed by continuing the signal from already transmitted areas into the area that is predicted. As shown in \cite{Dufaux1995, Ostermann2004}, temporal prediction mainly is obtained by performing motion compensation whereat for the region to be predicted a corresponding region in a previous frame is determined that matches the actual region best. In order to refer to this region a displacement vector, the so called motion vector, is transmitted as side information. Although, most encoders can switch between temporal and spatial prediction for different areas of a frame, no combined spatio-temporal prediction is performed. Up to now, only very few prediction methods exist that make use of spatial and temporal information at the same time. As two examples for spatio-temporal prediction, the ``Overlapped Block Motion Compensation'' by \cite{Orchard1994} or the ``Inter Frame Coding with Template Matching Spatio-Temporal Prediction'' by \cite{Sugimoto2004} should be mentioned.

We propose a new prediction method that operates in two stages in order to exploit temporal as well as spatial redundancies. In the first stage, we perform a preliminary motion compensated prediction for the signal in order to exploit the temporal correlations. Motion compensation normally provides a good estimate for the signal at the expense of only some small integer values defining the motion vector. As the motion compensation is performed only for the actually regarded area and independently from the surroundings of this area, the prediction signal may not fit into the surrounding area of the actual frame. Up to here, no use has been made of the spatial correlations. To cope with this circumstance, we propose a refinement step that generates a model of the signal based on the motion compensated predicted signal and the spatial neighborhood. The model generation takes the preliminary motion compensated area and the known surrounding area as input and aims at generating a new signal, the model, that incorporates the temporal as well as the spatial correlations. As we will show in Sec. \ref{sec:results} the produced model forms a better prediction for the actual signal than the pure motion compensated signal, leading to a significant reduction in data rate.


\section{Spatially refined prediction}

\begin{figure}
	\psfrag{m}[t][t][0.8]{$m$}%
	\psfrag{n}[t][t][0.8]{$n$}%
	\psfrag{x}[t][t][0.8]{$x$}%
	\psfrag{x0}[t][t][0.6]{$x_0$}%
	\psfrag{y}[t][t][0.8]{$y$}%
	\psfrag{y0}[t][t][0.6]{$y_0$}%
	\psfrag{t}[t][t][0.8]{$t$}%
	\psfrag{t0}[t][t][0.8]{$t=\tau$}%
	\psfrag{Agb}[l][l][0.8]{$\mathcal{A} = \mathcal{R} \cup \mathcal{B}$}%
	\psfrag{Block}[l][l][0.8]{$\mathcal{B}$}%
	\psfrag{Rgb}[l][l][0.8]{$\mathcal{R}$}%
	\psfrag{Pgb}[l][l][0.8]{$\mathcal{P}$}%
	\psfrag{Dgb}[l][l][0.8]{$\mathcal{D}$}%
	\centering
	\includegraphics[width=0.40\textwidth]{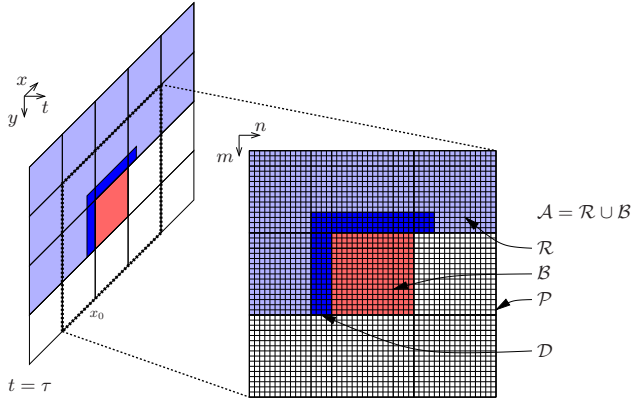}
	\caption{Projection area $\mathcal{P}$ with respect to the video sequence. $\mathcal{P}$ contains the approximation area $\mathcal{A}$, consisting of the area $\mathcal{R}$ of the reconstructed signal and the block $\mathcal{B}$ that will be predicted. Area $\mathcal{D}$, used for the implicit mode decision, is marked in dark blue.\vspace{-0.4cm}}
	\label{fig:prediction_area}
\end{figure}

We consider a block based prediction and coding scenario in which the blocks are processed in line scan order. In this context the ori\-ginal video sequence is depicted by $v\left[x,y,t\right]$ with the two spatial coordinates $x, y$ and the temporal coordinate $t$. Encoding and prediction is performed in frame $\tau$ for the block located at $x_0$ and $y_0$. We regard a new two dimensional coordinate system (see Fig. \ref{fig:prediction_area}) with spatial coordinates $m, n$ centered by the block $\mathcal{B}$ that will be predicted. The complete regarded area $\mathcal{P}$ is called projection area and is of size $M\times N$. It contains the block $B$, the area $\mathcal{R}$ of pixels from blocks that already have been transmitted and an area of padded samples, used for giving $\mathcal{P}$ a rectangular shape. Areas $\mathcal{R}$ and $\mathcal{B}$ together form the approximation area $\mathcal{A}$. Furthermore, area $\mathcal{R}$ contains area $\mathcal{D}$ used for the implicit mode decision that will be introduced later.

As mentioned above, the proposed algorithm works in two stages. First, a motion compensated prediction is performed for area $\mathcal{B}$. This is followed by a post processing step for refining the motion compensated prediction signal by means of spatial information contained in the already transmitted blocks. For the motion compensated prediction, in frame $\tau-1$ the block best matching the block $\mathcal{B}$ is determined using block matching. So only a displacement vector has to be transmitted in order to obtain a preliminary prediction.

The refinement step performs a two-dimensional selective approximation of the signal in $\mathcal{A}$ and is based on the selective extra\-polation from \cite{Kaup1998, Seiler2008}. It aims at generating a model of the signal in the whole area $\mathcal{P}$. Therefore, in area $\mathcal{A}$ the signal $s\left[m,n\right]$ is regarded as union of the motion compensated block in $\mathcal{B}$ and the already transmitted re-decoded blocks in $\mathcal{R}$. The re-decoded blocks in area $\mathcal{R}$ can be considered as a good estimate for the original signal $v\left[m,n\right]$, whereas the motion compensated block is only a preliminary estimate. Thus, we aim at refining the preliminary estimate by means of the good estimated surroundings. For that purpose, the regarded signal $s\left[m,n\right]$ is approximated by the parametric model $g\left[m,n\right]$ with
\begin{equation}
 g\left[m,n\right] = \sum_{\forall k \in \mathcal{K}} c_k \varphi_k \left[m,n\right]
\end{equation}
being the weighted superposition of two-dimensional basis functions $\varphi_k \left[m,n\right]$. In general, every set of two-dimensional basis functions can be used with the only constraint that the set has to be mutually orthogonal with respect to the projection area $\mathcal{P}$. The weighting factors $c_k$ are denoted as expansion coefficients and define the portion each basis function has of the original signal. The set $\mathcal{K}$ contains all basis functions used.

\begin{figure}
	\psfrag{x12}[t][t][0.8]{$0$}%
	\psfrag{x11}[t][t][0.8]{$10$}%
	\psfrag{x10}[t][t][0.8]{$20$}%
	\psfrag{x09}[t][t][0.8]{$30$}%
	\psfrag{x08}[t][t][0.8]{$40$}%
	\psfrag{x07}[t][t][0.8]{$50$}%
	\psfrag{v07}[r][r][0.8]{$0$}%
	\psfrag{v08}[r][r][0.8]{$10$}%
	\psfrag{v09}[r][r][0.8]{$20$}%
	\psfrag{v10}[r][r][0.8]{$30$}%
	\psfrag{v11}[r][r][0.8]{$40$}%
	\psfrag{v12}[r][r][0.8]{$50$}%
	\psfrag{z01}[r][r][0.8]{$0$}%
	\psfrag{z02}[r][r][0.8]{$0.1$}%
	\psfrag{z03}[r][r][0.8]{$0.2$}%
	\psfrag{z04}[r][r][0.8]{$0$}%
	\psfrag{z05}[r][r][0.8]{$0.1$}%
	\psfrag{z06}[r][r][0.8]{$0.2$}%
	\centering
	\includegraphics[width=0.35\textwidth]{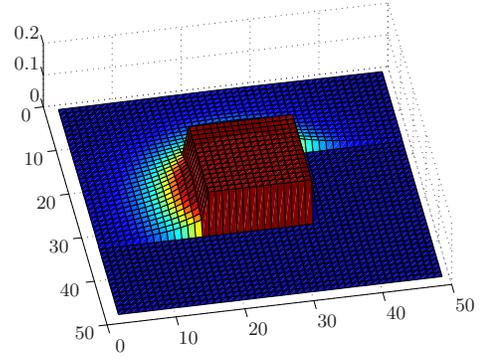}
	\caption{Example weighting function. The processed macroblock is weighted by $\mu = 0.25$, the surrounding is weighted according to an isotropic model with $\hat{\rho}=0.8$\vspace{-0.1cm}}
	\label{fig:weighting_function}
\end{figure}

The model is generated iteratively, according to \cite{Seiler2008}. There, in every iteration step one basis function is chosen to be added to the model and an estimate for its corresponding expansion coefficient is determined. In order to obtain this estimate, in every iteration step the residual between the original signal and the model is calculated. After this, weighted projections of the residual onto all basis functions are performed. Thereby the weighting function
\begin{equation}
 w \left[m,n\right] = \left\{ \begin{array}{ll} \rho\left[m,n\right] &,\ \forall \left(m,n\right) \in \mathcal{A} \\ 0 &,\ \mbox{else} \end{array} \right. 
\end{equation}
controls the influence pixels have on the approximation process depending on their position. Fig. \ref{fig:weighting_function} shows an example for a possible weighting function whereby the motion compensated block is weighted by $\mu = 0.25$ and the surrounding already transmitted blocks are weighted according to an isotropic model with $\hat{\rho} = 0.8$.
\begin{equation}
\rho \left[m,n\right] = \left\{ \begin{array}{ll} \hat{\rho}^{\sqrt{ \left(m-\frac{M-1}{2}\right)^2 + \left(n-\frac{N-1}{2}\right)^2}} &,\ \forall \left(m,n\right) \in \mathcal{R} \\ \mu &,\ \forall \left(m,n\right) \in \mathcal{B} \end{array} \right. 
\end{equation}
By using this weighting function all pixels from the motion compensated block in area $\mathcal{B}$ get the same weight. The weight for the pixels in the spatial neighborhood decreases with an increasing distance from the block to be predicted. Thus the pixels farther away from area $\mathcal{B}$ have less influence on the model generation than the closer ones. Generally, it is possible to control the influence the preliminary motion compensated prediction and the already transmitted spatial neighborhood have on the model generation by adjusting the parameters of the weighting function.

After finishing the iterations for generating the parametric mo\-del, area $\mathcal{B}$ is cut out from $g\left[m,n\right]$ and is used for final prediction of the actual block.

\begin{figure}
	\psfrag{x}[t][t][0.8]{$x$}%
	\psfrag{x0}[t][t][0.6]{$x_0$}%
	\psfrag{y}[t][t][0.8]{$y$}%
	\psfrag{y0}[t][t][0.6]{$y_0$}%
	\psfrag{t}[t][t][0.8]{$t$}%
	\psfrag{t0}[t][t][0.8]{$t=\tau$}%
	\psfrag{t1}[t][t][0.8]{$t=\tau-1$}%
	\centering
	\includegraphics[width=0.25\textwidth]{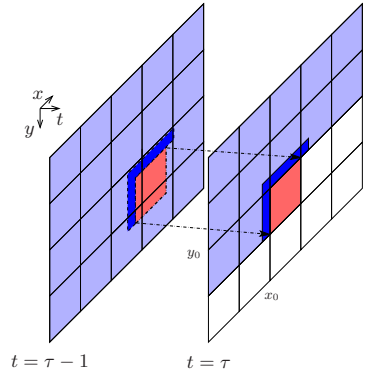}
	\caption{Motion compensated block $\mathcal{B}$ at $\left(x_0,y_0\right)$ and corresponding decision area $\mathcal{D}$ (dark blue) in frames $t=\tau$, respectively $t=\tau-1$.\vspace{-0.4cm}}
	\label{fig:decision_area}
\end{figure}

Unfortunately the refinement step cannot always result in better extrapolation results compared to the direct motion compensated prediction. Hence, we consider an implicit decision in order to determine if the the refinement step should be used or not. As the decision is performed on the decision area $\mathcal{D}$, the decision can be made without the need to transmit an additional bit for signaling which mode to use. The samples in the decision area already have been transmitted and thus are accessible at the receiver as well. For the mode decision they are compared with the parametric model $g\left[m,n\right]$ in this area on the one hand. On the other hand, they are compared with the corres\-ponding motion compensated pixels from frame $\tau -1$ denoted by $\hat{s}_{\mathrm{mc}} \left[m,n\right]$ (see Fig. \ref{fig:decision_area}). The decision criterion is the sum of absolute differences between $s\left[m,n\right]$ and $g\left[m,n\right]$ and between $s\left[m,n\right]$ and $\hat{s}_{\mathrm{mc}} \left[m,n\right]$ for all $\left(m,n\right) \in \mathcal{D}$. I.\ e.\ if 
\begin{equation}
 \sum_{\forall \left(m,n\right) \in \mathcal{D}} \hspace{-3mm} \left| s\left[m,n\right] - g\left[m,n\right] \right| <  \hspace{-3mm} \sum_{\forall \left(m,n\right) \in \mathcal{D}} \hspace{-3mm}  \left| s\left[m,n\right] - \hat{s}_{\mathrm{mc}} \left[m,n\right] \right|
\end{equation}
holds, the refined block is used for prediction, else the direct, motion compensated block is used. 

\begin{figure}
	\psfrag{Motion compensated block}[c][c][0.73]{Motion compensated block}
	\psfrag{Spatial neighborhood}[c][c][0.73]{Spatial neighborhood}
	\psfrag{Model generation}[c][c][0.73]{Model generation}
	\psfrag{Model quality evaluation}[c][c][0.73]{Model quality evaluation}
	\psfrag{Model}[c][c][0.73]{Model}
	\psfrag{is better than}[c][c][0.73]{is better than}
	\psfrag{motion compensated}[c][c][0.73]{motion compensated}
	\psfrag{prediction}[c][c][0.73]{prediction?}
	\psfrag{Cut out area B}[c][c][0.73]{Cut out area $\mathcal{B}$}
	\psfrag{out of gmn}[c][c][0.73]{out of $g\left[m,n\right]$}
	\psfrag{Use direct motion}[c][c][0.73]{Use direct motion}
	\psfrag{compensated block}[c][c][0.73]{compensated block}
	\psfrag{Predicted block}[c][c][0.73]{Predicted block}
	\psfrag{yes}[c][c][0.73]{yes}
	\psfrag{no}[c][c][0.73]{no}
	\psfrag{P}[c][c][0.73]{$\mathcal{P}$}
	\psfrag{B}[c][c][0.73]{$\mathcal{B}$}
	\psfrag{R}[c][c][0.73]{$\mathcal{R}$}
	\centering
	\includegraphics[width=0.44\textwidth]{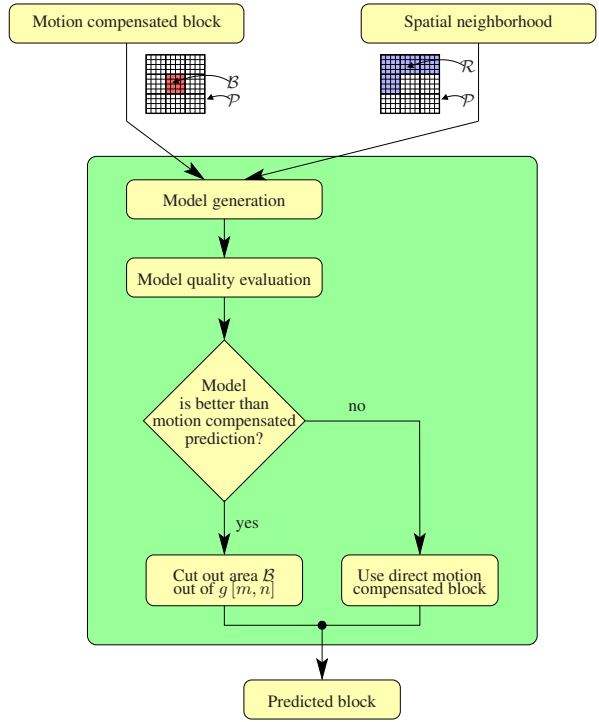}
	\caption{Block diagram: Spatio-temporal prediction by spatially refined motion compensation. \vspace{-0.4cm}}
	\label{fig:flowchart}
\end{figure}

In Fig. \ref{fig:flowchart} all steps for the spatio-temporal prediction by spatially refined motion compensation are shown in a block diagram.

\section{Simulation setup and results}\label{sec:results}

In order to illustrate the increased prediction quality obtainable by the spatial refinement of a motion compensated block compared to the unrefined pure temporal prediction, the algorithm described above is implemented in the H.264/AVC reference software JM10.2 with Baseline Profile, Level 2.0 \cite{Ostermann2004}. A subdivision of macroblocks in smaller blocks is not used, thus all operations are performed on macroblock level. Although the subdivision of macroblocks is not used in this simulation setup, a combination of the spatial refinement is also possible, as the refinement step can use any temporal predicted signal as input for further exploiting spatial correlations. The motion compensation is performed with quarter pixel accuracy, the search range is $16$ pixels, and one reference frame is used. To be able to directly compare the refined prediction with the unrefined one, the encoder's rate control is switched off and fixed QPs are used. Since the spatial refinement does not need any additional side information to be transmitted, it fits well into existing coding standards.

For evaluating the prediction quality, the first $99$ P-frames of the CIF-sequences ``Crew'', ``Flowergarden'', ``Foreman'', and ``Vimto'' are encoded at 30 frames per second. The 2D selective approximation performs $200$ iterations per macroblock for generating the model. As basis functions the functions of the two-dimensional Fourier transform are used since an efficient implementation in the transform domain exists \cite{Meisinger2004b}. Additionally, these basis functions are especially suited for natural images as it is possible to generate smooth areas as well as noise like areas and edges. For the transform into the Fourier domain a FFT of size $64\times64$ is applied.  The area $\mathcal{R}$ of the already transmitted pixels consists of the four neighboring macroblocks on top and to the left. The weighting function $w\left[m,n\right]$ emanates from an isotropic model with $\hat{\rho}=0.8$ and the factor $\mu=0.5$ for weighting the motion compensated block. Area $\mathcal{D}$ used for the implicit mode decision is a bar of $4$ pixels width that is located to the left and on top of the actual block (compare Fig. \ref{fig:prediction_area})

\begin{figure}
	\psfrag{s01}[t][t]{\color[rgb]{0,0,0}\setlength{\tabcolsep}{0pt}\begin{tabular}{c}$\mathrm{Rate} \ [\punit{kbit}/\punit{s}]$\end{tabular}}%
	\psfrag{s02}[b][b]{\color[rgb]{0,0,0}\setlength{\tabcolsep}{0pt}\begin{tabular}{c}$\PSNR \ [\punit{dB}]$\end{tabular}}%
	\psfrag{s04}[b][b]{}%
	\psfrag{s06}[][]{\color[rgb]{0,0,0}\setlength{\tabcolsep}{0pt}\begin{tabular}{c} \end{tabular}}%
	\psfrag{s07}[][]{\color[rgb]{0,0,0}\setlength{\tabcolsep}{0pt}\begin{tabular}{c} \end{tabular}}%
	\psfrag{s08}[l][l][0.68]{\color[rgb]{0,0,0}``Vimto'', refined}%
	\psfrag{s17}[l][l][0.68]{\color[rgb]{0,0,0}``Crew'', direct}%
	\psfrag{s18}[l][l][0.68]{\color[rgb]{0,0,0}``Crew'', refined}%
	\psfrag{s19}[l][l][0.68]{\color[rgb]{0,0,0}``Flowergarden'', direct}%
	\psfrag{s20}[l][l][0.68]{\color[rgb]{0,0,0}``Flowergarden'', refined}%
	\psfrag{s21}[l][l][0.68]{\color[rgb]{0,0,0}``Foreman'', direct}%
	\psfrag{s22}[l][l][0.68]{\color[rgb]{0,0,0}``Foreman'', refined}%
	\psfrag{s23}[l][l][0.68]{\color[rgb]{0,0,0}``Vimto'', direct}%
	\psfrag{s24}[l][l][0.68]{\color[rgb]{0,0,0}``Vimto'', refined}%
	\psfrag{x12}[t][t]{$0$}%
	\psfrag{x13}[t][t]{$500$}%
	\psfrag{x14}[t][t]{$1000$}%
	\psfrag{x15}[t][t]{$1500$}%
	\psfrag{x16}[t][t]{$2000$}%
	\psfrag{x17}[t][t]{$2500$}%
	\psfrag{x18}[t][t]{$3000$}%
	\psfrag{x19}[t][t]{$3500$}%
	\psfrag{v12}[r][r]{$25$}%
	\psfrag{v13}[r][r]{$30$}%
	\psfrag{v14}[r][r]{$35$}%
	\psfrag{v15}[r][r]{$40$}%
	\psfrag{v16}[r][r]{$45$}%

	\centering \vspace{-0.3cm}
	\includegraphics[height=6.5cm]{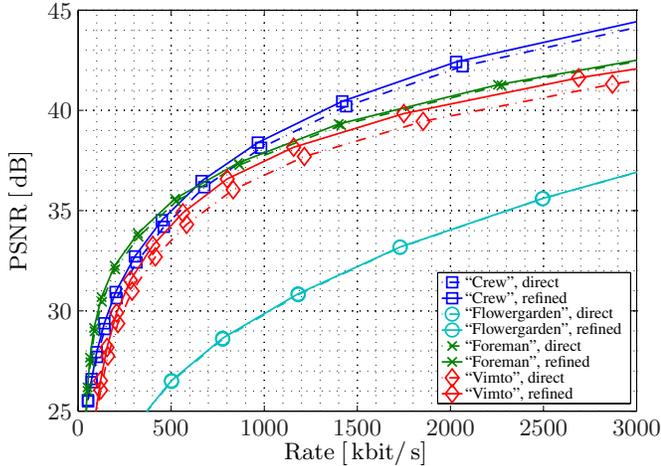}
	\caption{Rate-Distortion curves for the first $99$ P-frames of the used CIF-sequences at $30$ frames per second. Direct: H.264/AVC Baseline Profile, Level 2.0, one reference frame, fixed QPs. Refined:  model generation based on the direct motion compensated signal by 2D selective approximation with $200$ iterations, $\hat{\rho}=0.8$, and $\mu=0.5$\vspace{-0.4cm}}
	\label{fig:sim_results}
 
\end{figure}

Fig. \ref{fig:sim_results} shows the rate-distortion curves for the examined sequences in the range from $0$ to $3000$ $\punit{kbit}/\punit{s}$. The curves depicted by ``direct'' result from directly using the motion compensated block as prediction for the actual block, the ones depicted by ``refined'' result from applying the spatial refinement to the motion compensated block. Obviously, the spatial refinement leads to a better prediction compared to the direct motion compensation, resulting in less data rate needed to achieve the same $\punit{PSNR}$. For the sequence ``Flowergarden'' the gain is so small that it cannot be seen in Fig. \ref{fig:sim_results}. Therefore, Tab. \ref{tab:psnr_gain} summarizes for all sequences the maximum achievable gain in $\punit{PSNR}$ at a fixed data rate, respectively the maximum relative reduction in data rate for a fixed $\punit{PSNR}$. Additionally, the table shows the average relative rate reduction, calculated according to \cite{Bjontegaard2001}. Although no gain is visible for ``Flowergarden'' at the Rate-Distortion curves a small increase in prediction quality is possible. Regarding the other three sequences, the increased prediction quality caused by the spatial refinement becomes obvious and at maximum a relative reduction for the needed data rate of $14.79\%$ is obtainable. 

Regarding the examined sequences it becomes apparent that sequences such as ``Flowergarden'' with smooth translational motion do not profit very much by the spatial refinement. In such sequences, most of the blocks can be predicted very well using the temporal correlations only. On the other hand, sequences with heavy motion, occlusions or abrupt changes of the luminance gain a lot by incorporating spatial information for the prediction, since in these cases a pure temporal prediction only suboptimally fits the block to be coded.

\begin{table}
\small 
\centering
\begin{tabular}{|c|c|c|c|}
\hline
 & Max. $\PSNR$ & Max. Rate &  Avg. Rate \\
Sequence& Gain & Reduction  &  Reduction\\ \hline
``Crew'' & $0.45 \punit{dB}$ &  $9.57\%$  &  $6.60\%$ \\ \hline
``Flowergarden'' &$0.06 \punit{dB}$  & $1.02\%$ & $0.37\%$ \\ \hline
``Foreman'' &$0.21 \punit{dB}$  & $8.20\%$ & $3.84\%$ \\ \hline
``Vimto'' & $0.75 \punit{dB}$&$14.79\%$ &$11.22\%$ \\ \hline
\end{tabular}
\caption{Maximum achievable $\PSNR$ gain, maximum achievable relative rate reduction and average achievable relative rate reduction according to \cite{Bjontegaard2001}\vspace{-0.4cm}}
\label{tab:psnr_gain}
\end{table}

\begin{figure}
	\psfrag{s01}[t][t]{\color[rgb]{0,0,0}\setlength{\tabcolsep}{0pt}\begin{tabular}{c}$\mathrm{Rate} \ [\punit{kbit}/\punit{s}]$\end{tabular}}%
	\psfrag{s02}[b][b]{\color[rgb]{0,0,0}\setlength{\tabcolsep}{0pt}\begin{tabular}{c}$\PSNR \ [\punit{dB}]$\end{tabular}}%
	\psfrag{s05}[][]{\color[rgb]{0,0,0}\setlength{\tabcolsep}{0pt}\begin{tabular}{c} \end{tabular}}%
	\psfrag{s06}[][]{\color[rgb]{0,0,0}\setlength{\tabcolsep}{0pt}\begin{tabular}{c} \end{tabular}}%
	\psfrag{s07}[b][b]{}%
	\psfrag{s08}[l][l][0.68]{\color[rgb]{0,0,0}Refined MC + deblocking}%
	\psfrag{s17}[l][l][0.68]{\color[rgb]{0,0,0}Direct MC}%
	\psfrag{s18}[l][l][0.68]{\color[rgb]{0,0,0}Refined MC}%
	\psfrag{s19}[l][l][0.68]{\color[rgb]{0,0,0}Direct MC + deblocking}%
	\psfrag{s20}[l][l][0.68]{\color[rgb]{0,0,0}Refined MC + deblocking}%
	
	\psfrag{x12}[t][t]{$0$}%
	\psfrag{x13}[t][t]{$500$}%
	\psfrag{x14}[t][t]{$1000$}%
	\psfrag{x15}[t][t]{$1500$}%
	\psfrag{x16}[t][t]{$2000$}%
	\psfrag{x17}[t][t]{$2500$}%
	\psfrag{x18}[t][t]{$3000$}%
	\psfrag{x19}[t][t]{$3500$}%
	\psfrag{x20}[t][t]{$4000$}%
	\psfrag{x21}[t][t]{$4500$}%
	\psfrag{v12}[r][r]{$30$}%
	\psfrag{v13}[r][r]{$34$}%
	\psfrag{v14}[r][r]{$38$}%
	\psfrag{v15}[r][r]{$42$}%
	\psfrag{v16}[r][r]{$46$}%
	\centering \vspace{-0.3cm}

\includegraphics[height=6.5cm]{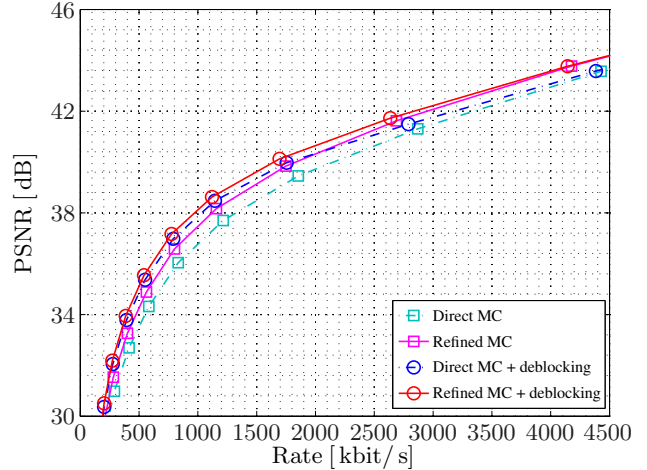}
	\caption{Rate-Distortion curves for the first $99$ P-frames of the CIF-sequence ``Vimto'' at $30$ frames per second. Comparison between the unrefined and refined motion compensation with and without H.264/AVC in-loop deblocking filter.\vspace{-0.4cm}}
	\label{fig:deblocking_compare}
\end{figure}

The simulations presented so far, had been carried out with H.264/AVC's in-loop deblocking filter switched off as a direct comparison between the motion compensated prediction and the refined prediction should be carried out. However, a common problem of all block based video codecs is their tendency to generate block artifacts. In order to compensate these artifacts H.264/AVC uses an in-loop deblocking filter \cite{List2003} that reduces visual artifacts as well as increases the coding efficiency. Since the main cause for blocking artifacts is the coarse quantization of the prediction error, the deblocking filter can also be combined with the spatially refined motion compensated prediction. In Fig. \ref{fig:deblocking_compare} the Rate-Distortion curves are illustrated for the first $99$ frames of the CIF-sequence ``Vimto'' at $30$ frames per second for the direct motion compensated prediction and the refined prediction, both with the deblocking filter switched on and off.  Apparently, for low data rates the refined motion compensated prediction used so far also suffers from blocking and is outperformed by the direct motion compensation with deblocking filter switched on. But by applying the deblocking filter on the video signal produced with the refined motion compensated prediction, the blocking artifacts can be reduced as well and the coding efficiency could be increased further. At high data rates, the degradation due to blocking only is weak, hence the curves with deblocking filter switched on and off converge and the existent reduction in data rate is generated by the spatially refined prediction. 


\section{Conclusion} \label{sec:conclusion}

As shown above, the proposed spatio-temporal prediction provides a very effective approach to enhance the prediction quality in video coding resulting in a reduced data rate needed for transmission of a sequence. The spatial refinement covers the lack of the motion compensated prediction to pay attention to the spatial correlations in a video sequence. Hence the proposed two-stage algorithm for prediction exploits temporal as well as spatial correlations in a video sequence. 

The implementation of the new prediction method into the H.264/AVC reference codec demonstrated that by exploiting spatial correlations in addition to temporal correlations the data rate can be reduced up to $14.79\%$ for the regarded sequences. As the idea of generating a model of the signal based on a preliminary prediction is a generic one, further investigations will focus on alternative methods for generating the signal model. 

Nevertheless, further work also will focus on combining the spatial refinement step with more sophisticated features of the H.264/AVC codec and on reducing the complexity of the spatial refinement so that it can be applied to smaller block sizes as well.


\end{document}